\documentclass[sigplan,10pt]{acmart}
\renewcommand\footnotetextcopyrightpermission[1]{}
\usepackage{xspace}
\newcommand{\sysname}{DeepCompile\xspace}

\AtBeginDocument{%
  }

\usepackage{amssymb}
\usepackage{pifont}
\newcommand{\xmark}{\ding{55}}
\usepackage{algorithm}
\usepackage{algorithmic}

\begin{document}

\title{DeepCompile: A Compiler-Driven Approach to Optimizing Distributed Deep Learning Training}

\author{Masahiro Tanaka}
\affiliation{
  \institution{Microsoft}
  \city{}
  \country{}
}
\email{mtanaka@microsoft.com}

\author{Du Li}
\affiliation{
  \institution{Microsoft}
  \city{}
  \country{}
}
\email{duli.personal@gmail.com}

\author{Umesh Chand}
\authornote{Work done while at Microsoft. Now at AMD.}
\affiliation{
  \institution{Microsoft}
  \city{}
  \country{}
}
\email{Umesh.Chand@amd.com}

\author{Ali Zafar}
\affiliation{
  \institution{University of Virginia}
  \city{}
  \country{}
}
\email{mzw2cu@virginia.edu}

\author{Haiying Shen}
\affiliation{
  \institution{University of Virginia}
  \city{}
  \country{}
}
\email{hshen@virginia.edu}

\author{Olatunji Ruwase}
\affiliation{
  \institution{Microsoft}
  \city{}
  \country{}
}
\email{olruwase@microsoft.com}

\renewcommand{\shortauthors}{Tanaka et al.}

\begin{abstract}
The rapid growth of deep learning models has increased the demand for efficient distributed training strategies. Fully sharded approaches like ZeRO-3 and FSDP partition model parameters across GPUs and apply optimizations such as prefetching and unsharding to reduce communication overhead. 
However, these systems lack fine-grained control over memory and communication scheduling, making it difficult to balance computation–communication overlap with memory requirements. 
Coordinating multiple optimizations such as prefetching and unsharding is also difficult, since their effects on memory usage can influence each other. To tackle these challenges, we propose \sysname, a compiler-based optimization framework for distributed training. \sysname transforms user-defined models into computation graphs and applies a series of profiling-guided optimization passes, each modifying the graph based on profiling information such as execution time and memory usage. This design allows each pass to flexibly insert, reorder, or remove operations such as all-gather and memory allocation, improving communication–computation overlap and reducing memory pressure. Each pass can access updated profiling feedback from earlier passes, enabling coordinated optimizations.
We further enhance \sysname by three additional optimizations: proactive prefetching, selective unsharding, and adaptive offloading. 
Our evaluation shows that \sysname achieves up to 1.28× and 1.54× speedups over ZeRO-3 and FSDP baselines, respectively, and up to a 7.01× throughput increase in settings with limited GPU resources using offloading.
\end{abstract}



\maketitle

\section{Introduction}\label{sec:introduction}

The rapid growth of deep learning has led to the emergence of increasingly large models. Modern architectures often contain tens or even hundreds of billions of parameters~\cite{brown2020language,zhang2022opt,abdin2024phi3technicalreporthighly,grattafiori2024llama3herdmodels,chowdhery2022palm}, resulting in immense computational demands. While deep learning compilers~\cite{chen2018tvm,xla,pytorch2} have been highly successful in improving training efficiency on a single accelerator device, such as a GPU, efficient parallelization across multiple devices has become essential to fit large models into memory.

There are several parallelization strategies for training large models. In data parallelism, all model parameters are replicated across multiple GPUs. This approach is natively supported in many deep learning frameworks, including PyTorch~\cite{paszke2019pytorch}. 
Pipeline parallelism~\cite{huang2019gpipe} instead splits the model by layers, assigning each block to a different GPU. Tensor parallelism~\cite{shoeybi2019megatron} divides parameter tensors within a layer across GPUs, where each device computes a partial result that is later aggregated.
In addition to these, the {\em fully sharded approach}, which is implemented in DeepSpeed ZeRO-3~\cite{rajbhandari2020deepspeed} (hereafter ZeRO-3) and Fully Sharded Data Parallel (FSDP)~\cite{fsdp_paper}, partitions the parameters of each layer across multiple GPUs and gathers them through {\em all-gather} communication before each layer’s computation begins.

Although existing parallelization strategies have enabled large-scale model training, two key challenges remain unsolved:

\noindent
\textbf{Challenge 1: Fine-grained control over communication and memory scheduling.}  Existing systems often lack the ability to precisely control when to initiate communication or allocate and release memory buffers. As a result, it is difficult to optimize the overlap between communication and computation, or to efficiently manage memory usage that fluctuates during execution. 

\noindent
\textbf{Challenge 2: Coordinating multiple optimizations.} As existing deep learning compilers apply a sequence of independent optimization passes, such as kernel fusions and operator reordering, distributed training also requires independent optimizations such as communication scheduling and memory planning. However, there is currently no unified mechanism to combine these optimizations and handle their interactions effectively.

To illustrate Challenge 1, consider {\em prefetching} in the fully sharded approach. Initiating all-gather communication earlier can improve overlap with computation, but prefetching too much data can exhaust available memory, as the buffers cannot be released until all dependent computations are completed.

A representative case of Challenge 2 involves combining prefetching with {\em unsharding}, which retains as many parameters in their unsharded form as possible to reduce communication overhead. To combine these two optimizations effectively, the system must accurately estimate how much memory remains available after prefetching is applied.

Existing systems partially address these challenges. Runtime-based approaches such as ZeRO-3 and FSDP support optimizations like prefetching and unsharding, but lack fine-grained control over the timing of communication and memory scheduling in response to runtime memory changes. Compiler-based systems such as Alpa~\cite{mo2022alpa}, Unity~\cite{unger2022unity}, nnScalar~\cite{lin2024nnscaler} automatically combine different parallelization strategies, but do not adapt to dynamic memory behavior during execution.

To bridge this gap, we introduce \sysname, a compiler-based optimization framework for distributed training. \sysname uses an existing compiler to convert a model into a computation graph and applies a series of transformations to optimize multi-GPU execution. These transformations improve communication overlap and reduce memory pressure based on profiling information that captures execution time and memory usage trends across the forward and backward passes.

This approach addresses the two key challenges discussed earlier:

\noindent
{\bfseries Graph transformations enabling fine-grained memory-aware scheduling.}  
By directly transforming the computation graph, \sysname can flexibly inject, reorder, and remove operations such as all-gather. This fine-grained control enables precise placement of communication and memory operations, which is difficult to express in user-level code. Furthermore, by capturing operator-level memory usage patterns across forward and backward passes, \sysname identifies optimal points to initiate communication and allocate or release memory based on runtime memory availability.

\noindent
{\bfseries Composable and unified optimization passes with profiling feedback.}
\sysname organizes optimizations into modular \emph{optimization passes}, each transforming the computation graph based on profiling data such as execution time and memory usage. While each pass can operate sequentially, later passes can access updated profiling data from the transformed graph to observe and respond to the effects of earlier passes. This design enables multiple distributed training optimizations to be composed in a globally coordinated manner.

To evaluate the effectiveness of this design, we implemented the fully sharded approach in \sysname as an optimization pass, similar in functionality to ZeRO-3~\cite{rajbhandari2020deepspeed} and FSDP~\cite{fsdp_paper} (\S\ref{subsec:zero3_sharding}). This pass demonstrates that the fully sharded paradigm can be captured within the optimization pass framework, and serves as a foundation for building additional compiler-based optimizations.

Building on this foundation, we design and incorporate three additional optimization passes into \sysname. Each pass targets a distinct aspect of communication or memory optimization in fully sharded training.

\begin{itemize}
\item \textbf{Proactive prefetching (\S\ref{subsec:prefetch})} initiates all-gather operations earlier to improve overlap between communication and computation, while taking into account dynamic memory availability during execution.
\item \textbf{Selective unsharding (\S\ref{subsec:unsharding})} avoids sharding for selected parameters based on operator-level memory profiling, reducing all-gather operations for sharded parameters and adapting to memory usage trends.
\item \textbf{Adaptive offloading (\S\ref{subsec:adaptive_offload})} transfers optimizer states, such as the momentum and variance buffers used by the Adam optimizer~\cite{kingma2014adam}, to CPU memory when GPU memory is insufficient. These transfers are scheduled to overlap with computation to hide latency.
\end{itemize}

All of these passes are applied after the optimization pass enabling the baseline fully sharded functionality described in \S\ref{subsec:zero3_sharding}. Moreover, as we show in later experiments, combining proactive prefetching and selective unsharding yields significant performance improvements over applying either pass in isolation.

SimpleFSDP~\cite{zhang2024simplefsdpsimplerfullysharded} takes a similar approach to \sysname by using compilation to improve the placement of all-gather operations in the fully sharded approach. However, it is specialized for FSDP and focuses solely on prefetching, along with fused all-gather communications. As a result, it does not support composing multiple modular optimization passes. Coordination across passes using profiling feedback—for example, adapting later optimizations to the effects of earlier ones—is not considered in its design.

We evaluated \sysname on large-scale distributed training tasks using Llama 3 70B and Mixtral 8x7B MoE models. Across all configurations, \sysname consistently improved training efficiency compared to baselines, including ZeRO-3, FSDP, and setups using PyTorch compiler optimizations. It achieved up to 1.28× improvement on Llama 3 70B and up to 1.54× on the Mixtral 8x7B MoE model. Additionally, it delivered a 7.01× increase in throughput in low-GPU settings where, even with parameter partitioning, the model does not fit in GPU memory without offloading.

\section{Background and Motivating Example}\label{sec:background}

The rapid scaling of Transformer-based architectures~\cite{vaswani2017attention} has driven a dramatic increase in model size and computational demands. While accelerators like GPUs are standard, training state-of-the-art models now requires large-scale distributed training. For instance, Llama-3~\cite{grattafiori2024llama3herdmodels} was trained on 16,384 NVIDIA H100 GPUs over 54 days. Even fine-tuning tasks frequently span dozens of GPUs. As such, reducing overhead from inter-device communication and memory usage has become essential for efficient training.

\subsection{Parallelization Strategies}

To address the computational and memory demands of large-scale models, several parallelization strategies have been developed.

\textit{Data parallelism} replicates the entire model on each GPU and splits the input data. While easy to adopt, it cannot scale to very large models that exceed single-GPU memory.
\textit{Pipeline parallelism} splits the model by layers into sequential stages across GPUs. It reduces memory usage but requires careful balancing of stage workloads and often manual tuning.
\textit{Tensor parallelism} partitions individual weight matrices across devices and aggregates partial results using all-reduce. Though effective for standard architectures with well-supported implementations (e.g., GPT~\cite{brown2020language}), it often requires non-trivial engineering to adapt to variants such as multi-query attention~\cite{shazeer2019fast} or grouped-query attention~\cite{ainslie2023gqa}.

An alternative to pipeline and tensor parallelism is the \textit{fully sharded approach}, which partitions the parameters of each layer across multiple GPUs. Frameworks such as DeepSpeed ZeRO-3~\cite{rajbhandari2020deepspeed} and Fully Sharded Data Parallel (FSDP)~\cite{fsdp_paper} adopt this strategy. In this approach, each GPU gathers its required parameter shards using all-gather communication before computing each layer. To reduce peak memory usage, the gathered parameters are discarded immediately after the layer’s computation finishes. 
This approach is widely adopted because it can be applied to arbitrary model architectures without requiring manual restructuring. In this work, we focus on optimizing this fully sharded approach.

\subsection{Deep Learning Compiler}

Deep learning compilers have rapidly advanced in recent years~\cite{chen2018tvm,xla,pytorch2}. They translate user-defined models into computation graphs, which enables graph-level analysis and flexible optimizations such as operator reordering, fusion, and memory reuse. 

Major frameworks have been integrating compiler capabilities into their ecosystems. For example, PyTorch now includes its own compiler infrastructure to enable graph-based optimizations~\cite{pytorch2}. Since these compilers can be applied to the vast number of existing model implementations, they are now widely used in practice.

\subsection{Motivating Example}

The fully sharded approach, as implemented in systems like ZeRO-3 and FSDP, employs runtime optimizations such as prefetching and unsharding. However, these optimizations alone are not sufficient to fully improve training efficiency.

Let us take prefetching as an example. Prefetching aims to reduce communication overhead by initiating all-gather operations earlier than the layer where the parameters are actually needed, thereby overlapping communication with computation.
Because the fully sharded approach allocates a large buffer to store the gathered parameters, initiating all-gather earlier extends the buffer’s lifetime and increases memory pressure. Therefore, prefetching needs to be scheduled based on memory usage patterns throughout the forward and backward passes.

\begin{figure}
\begin{center}
\includegraphics[scale=0.40]{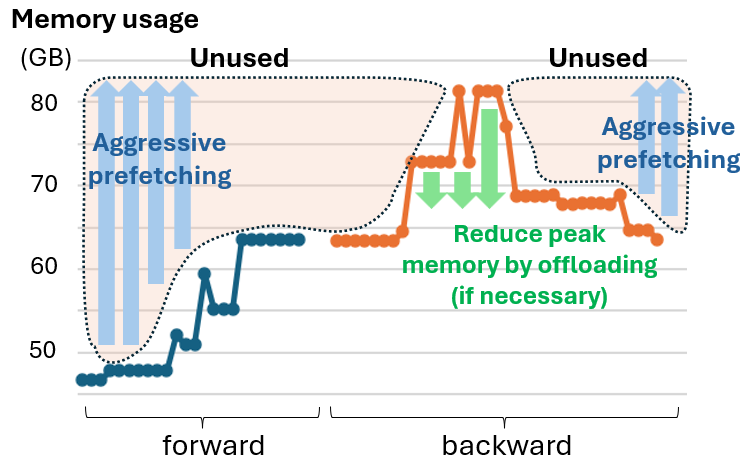}
\end{center}
\caption{Memory usage trends and scheduling opportunities for prefetching and offloading. Profile of several final layers with a sequence length of 4096 and a vocabulary size of 128k. Significant memory spikes are observed in the log-softmax and negative log-likelihood loss layers in the backward pass.}
\label{fig:mem_example}
\end{figure}

Figure~\ref{fig:mem_example} illustrates a typical pattern of GPU memory usage during forward and backward passes. Memory usage gradually increases during the forward pass as activations are stored for the following backward pass, and decreases during the backward pass as activations are released. At the beginning of the forward pass, a substantial amount of unused memory is available, providing an opportunity to prefetch aggressively and increase computation-communication overlap. In contrast, near the end of the forward pass and throughout the backward pass, memory usage is high, limiting the opportunity for prefetching due to reduced memory availability. As the backward pass progresses and memory is gradually freed, more aggressive prefetching becomes possible again.
Existing systems such as ZeRO-3 and FSDP typically allow users to specify a fixed buffer size for prefetching. However, this static configuration cannot adapt to changes in available memory as computation progresses, limiting the potential benefits of prefetching.

The same issue arises with offloading, which transfers data such as optimizer states (e.g., momentum and variance in the Adam optimizer~\cite{kingma2014adam}) to host memory to reduce GPU memory usage during forward and backward passes. This is possible because optimizer states are only needed after the backward pass, when updating parameters.
Though offloading can significantly reduce peak GPU memory usage, the associated data transfers introduce overhead. A naive strategy is to offload optimizer states before the forward pass and reload them after the backward pass finishes, since they are not used during computation. However, this leads to idle GPU time during data transfers.

Instead, data transfers and computation can be overlapped. By initiating transfers at the beginning of the forward pass and synchronizing their completion just before memory usage reaches its peak, it is possible to hide the cost of data movement. In the backward pass, the opposite strategy can be applied: as memory usage decreases, transfers from host memory to GPU can be initiated. To achieve this, the system must accurately track how memory usage changes during execution and place offload, reload, and synchronization operations at appropriate points.

These observations motivate the design of \sysname, a compiler-based approach for systematically optimizing fully sharded training.

\section{System Design}
\label{sec:design}

We now describe the design of \sysname, a compiler-based system that enables graph-level transformations to support flexible and coordinated optimization of distributed training.

\begin{figure}
\begin{center}
\includegraphics[scale=0.45]{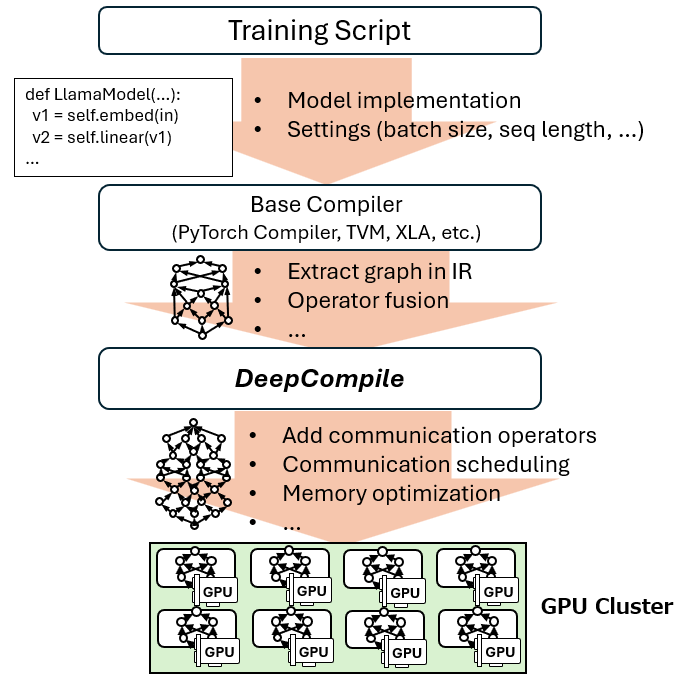}
\end{center}
\caption{Workflow of \sysname. \sysname transforms the IR generated by a base compiler, injecting distributed training optimizations before execution on GPUs.}
\label{fig:workflow}
\end{figure}

Figure~\ref{fig:workflow} shows the overall workflow. Starting from a user-defined training script and a model implementation written in a framework such as PyTorch, a base compiler (e.g., the PyTorch compiler) lowers the model into an intermediate representation (IR) as a computation graph. \sysname takes this graph as input, transforms it by adding communication operations for distributed training, and applies optimizations. The resulting graph is then deployed to a runtime engine running on GPU servers.

Note that \sysname does not assume that distributed training logic is manually written in the model code. Instead, it targets standard, framework-native implementations, such as those commonly found on the HuggingFace Model Hub\footnote{\url{https://huggingface.co/models}}, and programmatically adds necessary operators including communication and synchronization operations to the graph.

To support dynamic and memory-aware optimizations, \sysname organizes its graph transformations as a sequence of {\em optimization passes}, as is common in existing deep learning compilers.
Each pass rewrites the computation graph based on a specific strategy, such as inserting communication operators or reordering memory-intensive computations. After applying each pass, \sysname executes the modified graph to collect runtime profiling data, including operator execution times, communication overhead, and memory usage trends. This information is then used to guide the next optimization pass. By repeating this process, \sysname incrementally refines the graph with awareness of runtime behavior.

In addition, to account for memory usage changes that arise during actual training, \sysname periodically runs short training iterations between groups of optimization passes. While \sysname focuses primarily on the forward and backward passes, other parts of the training process, such as parameter updates, can significantly affect memory usage. For example, the Adam optimizer typically allocates a large buffer after the first backward pass completes. To capture such changes, \sysname runs several training iterations to reflect the updated memory dynamics, then applies another round of optimization passes adapted to the new conditions.

\begin{figure}
\begin{center}
\includegraphics[scale=0.2]{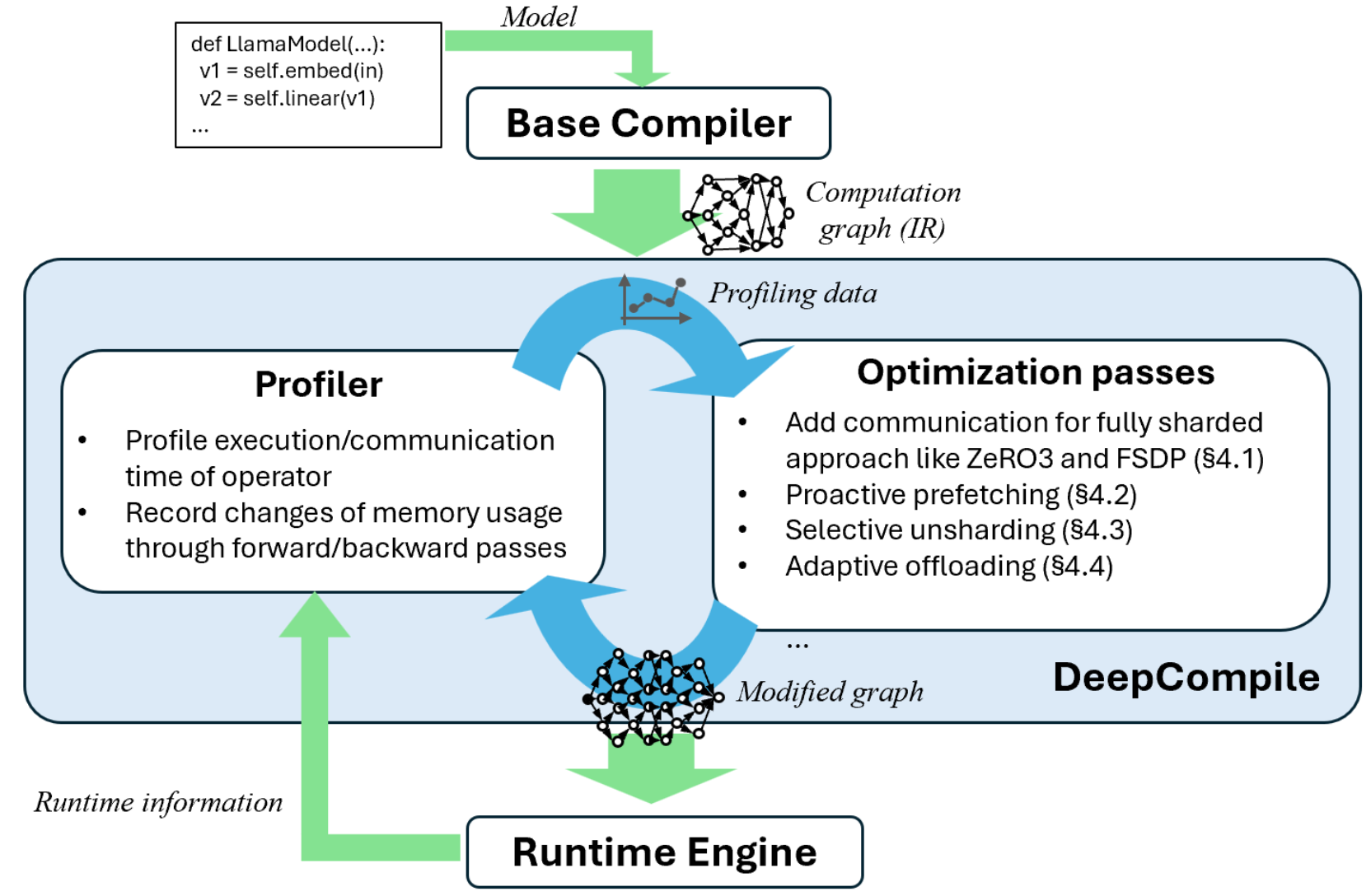}
\end{center}
\caption{Profiling-guided optimization loop in \sysname. \sysname applies optimizations in iterative phases, with periodic training to reflect memory dynamics and coordinate passes such as prefetching and offloading.}
\label{fig:opt_loop}
\end{figure}

Figure~\ref{fig:opt_loop} illustrates this two-level loop: an inner loop of optimization and profiling for each optimization pass, and an outer loop that periodically runs training to reflect changes in the runtime environment. By alternating between these loops, \sysname adapts its transformations to the evolving memory and execution characteristics of the model. This enables the coordination of multiple optimizations such as prefetching, unsharding, and offloading in a unified and profile-guided manner.

While \sysname is applicable to a wide range of optimization problems, this paper focuses on one representative use case: the fully sharded approach, as implemented in ZeRO-3 and FSDP. We take advantage of \sysname’s graph-level transformation and profiling infrastructure to implement this use case. Specifically, we incorporate a series of optimization passes that target key performance bottlenecks, including proactive prefetching to overlap communication with computation, selective unsharding to reduce redundant data communication, and adaptive offloading to mitigate GPU memory pressure.

In the following sections, we describe the implementation of this fully sharded training strategy using \sysname and detail the design and impact of each optimization pass.



\section{Optimizations}\label{sec:optimizations}

\sysname enables a broad range of optimizations by leveraging a computational graph extracted by the base compiler. In this paper, we begin by describing how the fully sharded approach is expressed within our framework, and then introduce additional optimizations that reduce communication overhead and memory requirements.

\subsection{Fully-sharded approach}\label{subsec:zero3_sharding}

In the fully sharded approach, each parameter tensor is evenly partitioned across all GPUs. Before a layer is computed, each GPU gathers the required parameter shards via an all-gather communication, reconstructing the full parameters locally. Once the layer computation is complete, the gathered parameters are discarded to free memory.

\begin{figure}
\begin{center}
\includegraphics[scale=0.16]{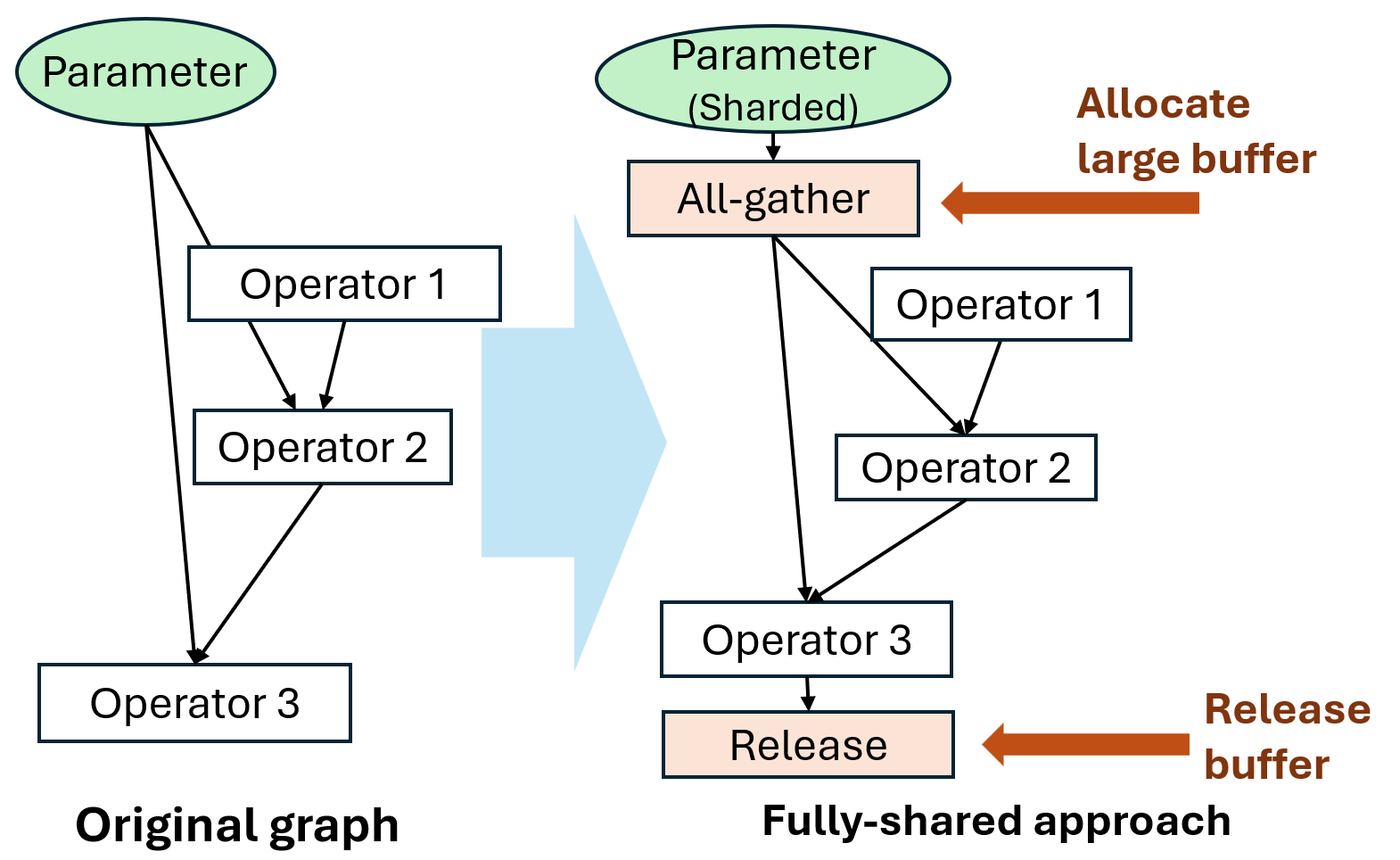}
\end{center}
\caption{Inserting all-gather and release operations for the fully sharded approach. This minimizes buffer lifetime by aligning communication and memory operations with actual usage.}
\label{fig:graph_mod}
\end{figure}

In existing implementations of the fully sharded approach, such as ZeRO-3 and FSDP, communication and memory operations are typically triggered by hooks inserted at layer boundaries in the Python module structure. These hooks invoke {\em all-gather} operations and manage memory allocation and release in response to the forward and backward pass.

In contrast, \sysname operates by modifying the computation graph to insert all-gather and release operations, similar to existing frameworks. However, unlike systems that rely on layer boundaries or manually defined blocks, \sysname can flexibly inject or reorder such communication and memory operations based on the dependencies among operators.

As a first step toward enabling the fully sharded approach, \sysname schedules all-gather operations just before each parameter's first use, and release operations immediately after its last use, minimizing the lifetime of each buffer. Figure~\ref{fig:graph_mod} shows an example of this process. If a parameter is used by multiple operators, \sysname analyzes their dependencies to ensure that the buffer is released only after its final use.

Although this scheduling improves memory efficiency, it does not necessarily improve execution speed. To address this, \sysname applies a series of optimization passes. The initial pass ensures that the model fits within available memory, while subsequent passes improve training efficiency by adjusting the timing of communication, avoiding unnecessary sharding, and overlapping data transfers with computation.

\subsection{Proactive prefetching}\label{subsec:prefetch}

\begin{figure}
\begin{center}
\includegraphics[scale=0.3]{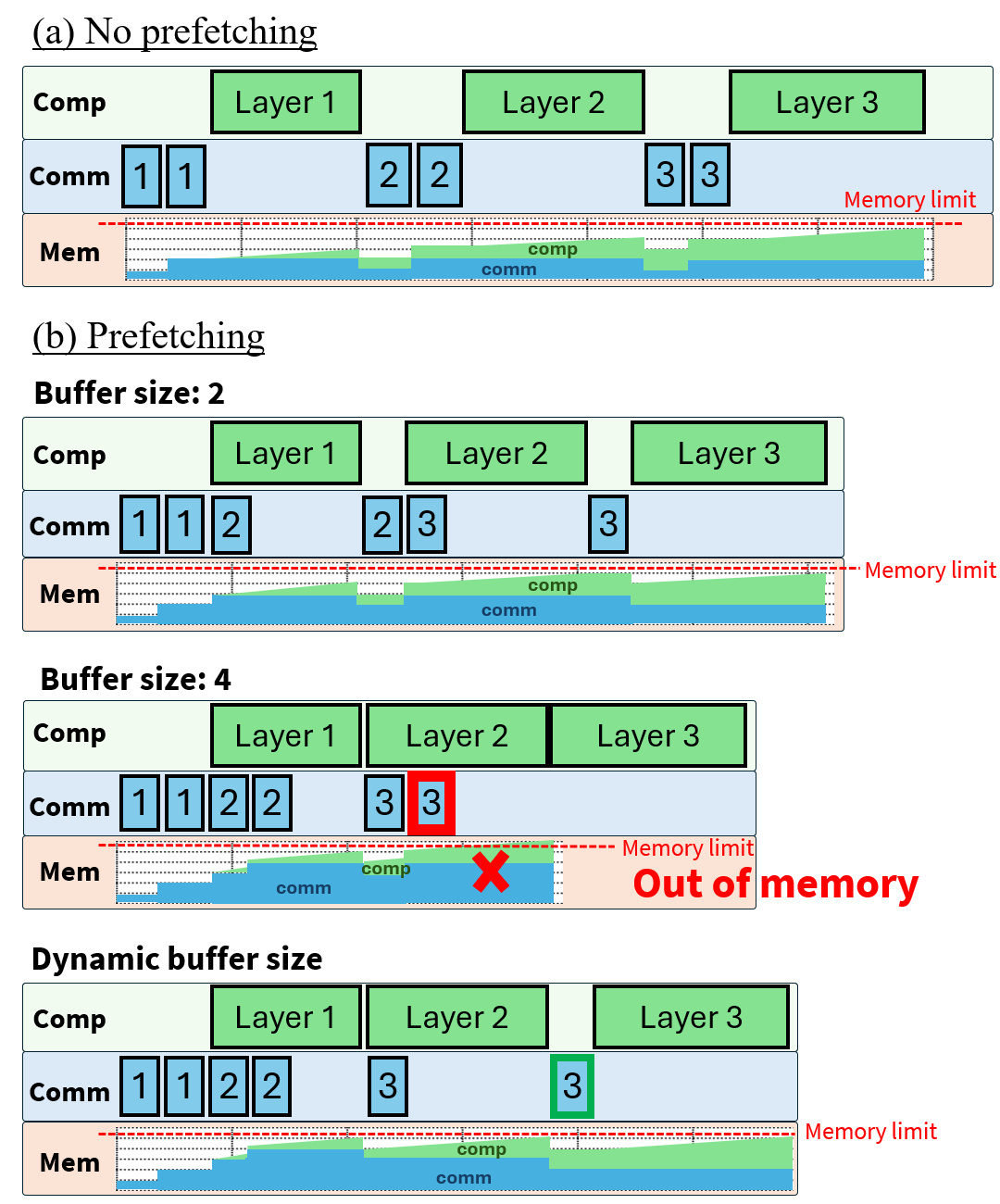}
\end{center}
\caption{Impact of prefetching on memory usage and communication–computation overlap.
(a) Without prefetching, all-gather operations are issued just before computation, resulting in no overlap. (b) Prefetching enables overlap, but fixed buffer sizes may cause out-of-memory errors. A dynamic strategy adapts to memory availability and avoids such failures.}
\label{fig:prefetch}
\end{figure}

The initial graph transformation for the fully sharded approach in \sysname produces behavior similar to that of ZeRO-3 and FSDP, where each layer triggers an all-gather before its computation and releases the gathered parameters immediately afterward. Figure~\ref{fig:prefetch}(a) illustrates this baseline behavior during a forward pass. Computation (green boxes) and all-gather communication (blue boxes) are executed sequentially per layer, resulting in no overlap.

The bottom track in each timeline shows memory usage over time. Activation memory (green), used for intermediate computation, accumulates across layers and is retained until the backward pass. In contrast, memory used for all-gather communication (blue) increases when the operation begins and is released once the corresponding computation completes. As a result, more memory is typically available earlier in the forward pass and later in the backward pass.

To reduce communication overhead, systems such as ZeRO-3 and FSDP support prefetching, which launches all-gather operations earlier than strictly required to enable communication–computation overlap. Figure~\ref{fig:prefetch}(b) demonstrates how different prefetching strategies impact memory usage and execution behavior. With a fixed prefetching buffer size of 2, safe overlap is achieved without exceeding memory limits. However, a larger buffer size of 4 causes an out-of-memory error due to increased memory pressure from overlapping communication and accumulating activation memory.

To prevent such failures, ZeRO-3 and FSDP employ a static buffer size and limits prefetching to the parameters that fit within that bound. While safe, this conservative strategy fails to adapt to dynamic changes in memory availability during training. As illustrated in Figure~\ref{fig:mem_example}, memory usage fluctuates significantly throughout execution. Fine-grained control is required to determine exactly when to initiate each all-gather operation. The bottom timeline in Figure~\ref{fig:prefetch}(b) exemplifies such dynamic scheduling, which is not feasible in existing systems like ZeRO-3 or FSDP.

To address this limitation, we introduce an optimization pass for {\em proactive prefetching}, which schedules all-gather operations as early as possible while ensuring that memory constraints are not violated. By profiling memory usage and all-gather buffer sizes after applying the optimization pass for the fully sharded approach, \sysname can estimate the total memory footprint of earlier communication.

\begin{figure}
\begin{center}
\includegraphics[scale=0.3]{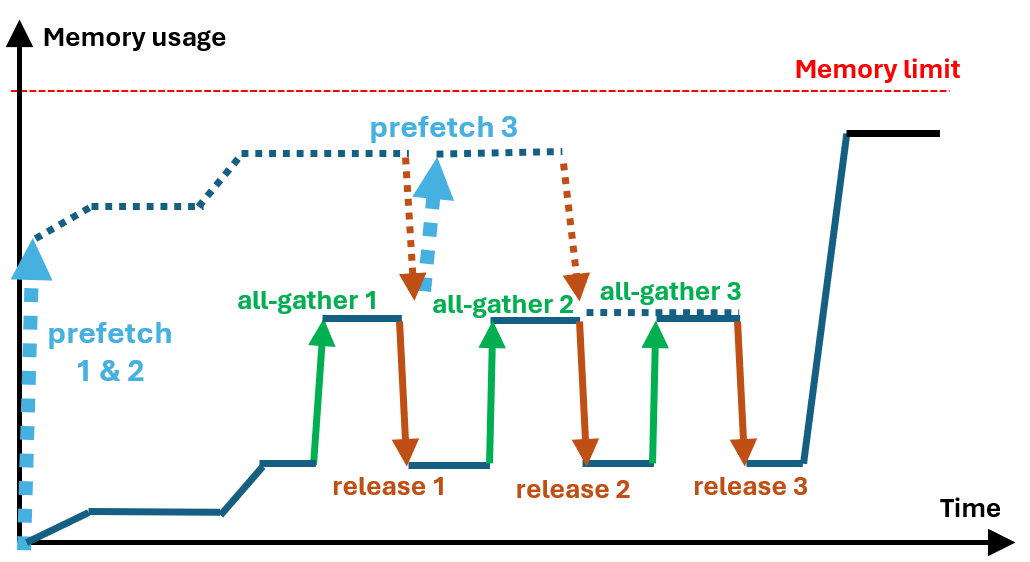}
\end{center}
\caption{Memory usage and prefetch scheduling decisions. Solid lines indicate memory usage without prefetching; dotted lines show usage with proactive prefetching. All-gather and release operations are labeled accordingly.}
\label{fig:prefetch_plan}
\end{figure}

Figure~\ref{fig:prefetch_plan} illustrates how this scheduling works. Solid lines indicate memory usage without prefetching, while dotted lines show usage with proactive prefetching. In this example, three all-gather operations are scheduled. Based on the memory budget, the first two can be safely prefetched at the beginning. The third, however, would exceed the memory limit if issued immediately, so it is delayed until after the buffer for the first parameter has been released. This approach maximizes prefetching opportunities while maintaining safe memory usage throughout execution.

\begin{table}[t]
\caption{Notations}
\label{tab:notation}
\centering
\begin{tabular}{ll}
\toprule
\textbf{Symbol} & \textbf{Description} \\
\midrule
$S_0$ & Initial schedule of operators $[o_1, o_2, \ldots, o_N]$ \\
$M$ & Total memory usage limit \\
$M_{\text{prefetch}}$ & Size limit for the buffer size of a prefetch group \\
$N$ & Number of operators \\
$OS$ & Fragments of optimizer states $[os_{1}, os_{2}, \ldots]$ \\
$\text{P}_{\text{mem}}(o)$ & Profiled memory usage before operator $o$ \\
$\text{B}_{\text{ag}}(o)$ & Buffer size allocated by all-gather operator $o$ \\
$\text{B}_{\text{os}}(os_i)$ & Size of optimizer state fragment $os_i$ \\
$Fuse(U)$ & Produce operator fusing all-gathers in $U$ \\
$T_c(V)$ & Communication time for $V$ \\
$\alpha$ & Fusion threshold parameter\\
\bottomrule
\end{tabular}
\end{table}

\begin{algorithm}[t]
\caption{Proactive prefetching}\label{alg:prefetch}
\begin{algorithmic}[1]
\REQUIRE Initial schedule $S_0$, Memory limit $M$, Size limit for prefetch group $M_{\text{prefetch}}$.

\STATE $S \gets []$ \COMMENT{Operators in new schedule}
\STATE $U \gets []$ \COMMENT{Unscheduled all-gathers}
\FOR{$i = N$ \TO $2$}
    \IF{$o_i$ is \textit{allgather}}
      \STATE $\Tilde{m}_U \gets \sum_{o \in U \cup \{o_i\}} \text{B}_{\text{ag}}(o)$
      \STATE $\Tilde{m}_{i-1} \gets \text{P}_{\text{mem}}(o_{i-1}) + \Tilde{m}_U$
      \IF{$\Tilde{m}_{i-1} < M$ and $\Tilde{m}_U < M_{\text{prefetch}}$}
        \STATE $U \gets U \cup \{o_i\}$ \COMMENT{Add $o_i$ to prefetch list}
      \ELSE
        \STATE $S_f \gets \text{Fuse}(U)$, $U \gets []$
        \STATE Append $S_f$ to $S$ \COMMENT{Schedule fused allgathers}
    \ENDIF
  \ELSE
    \STATE Append $o_i$ to $S$
  \ENDIF
\ENDFOR
\STATE $S_f \gets \text{Fuse}(U)$
\STATE Append $S_f$ to $S$ \COMMENT{Schedule all remaining allgathers}
\RETURN $S$
\end{algorithmic}
\end{algorithm}

Algorithm~\ref{alg:prefetch} formalizes this procedure.
It iterates over the operators in the initial schedule $S_0$ in reverse order. The schedule $S_0$ is produced by the preceding optimization pass that enables the fully sharded approach.
For each operator $o_i$, if it is an all-gather operation, the algorithm checks whether it can be moved earlier in the schedule without violating the memory constraint. This decision is based on the profiled memory usage and the required buffer size.

If the estimated memory usage remains below the limit, the all-gather operator $o_i$ is added to a group $U$ of unscheduled all-gather operators that can be moved earlier. If the usage exceeds the limit, the algorithm schedules the operators in $U$ at the current position, applying fusion based on the \texttt{Fuse} function. This strategy maintains memory usage within the limit $M$.

We also set the limit $M_{\text{prefetch}}$ on the prefetch group, as moving too many all-gather operations earlier does not provide any additional benefit. By imposing this limit, we ensure that some memory remains available for selective unsharding, which is discussed in the next section.

The \texttt{Fuse} function decides whether to fuse a set of all-gather operators and returns a sequence of fused or unfused operations.
For small message sizes, communication time increases slowly and remains nearly flat. As a result, combining multiple small all-gather operations into a single larger one can significantly reduce communication overhead.
The $\text{Fuse}(U)$ operation fuses two all-gather calls if the following condition holds: if \( T_c(V_1) + T_c(V_2) > \alpha \cdot T_c(V_1 + V_2) \), where \( V_1 \) and \( V_2 \) are the data sizes of two operations. Here, $T_c(V)$ is profiled communication time for size $V$, and \( \alpha \) is a tunable parameter.

SimpleFSDP~\cite{zhang2024simplefsdpsimplerfullysharded} takes a similar approach to ours to optimize the fully sharded approach using a compiled computation graph. It performs memory-aware fusion of all-gather operations to reduce communication overhead. However, SimpleFSDP does not consider how memory usage dynamically changes during the forward and backward passes to determine how early each all-gather can be issued. In contrast, our proactive prefetching algorithm uses profiling data to make this decision, allowing communication to be overlapped with computation while staying within memory constraints.

\subsection{Selective unsharding}\label{subsec:unsharding}

After applying the optimization pass for proactive prefetching, some free memory may remain.
Selective unsharding leverages this remaining memory by keeping certain parameters in their unsharded form after they are gathered and delaying their release until the parameter update step, which occurs after the backward pass.
By avoiding repeated all-gather operations for these parameters, this approach reduces communication overhead.

This technique is especially beneficial when used with gradient accumulation, a common method to increase the effective batch size without increasing GPU memory usage. Given a gradient accumulation step of $n$, the model performs $n$ forward and backward passes while accumulating gradients before updating parameters. Since no parameter updates occur during this period, the gathered parameters can remain unsharded across multiple forward and backward passes. After a backward pass, the optimizer states are reloaded to GPU memory to update parameters.

We run profiling after applying proactive prefetching to measure the peak memory usage. Then, we select as many parameters to keep unsharded as possible, ensuring that the total buffer size does not exceed the memory limit.
The selection considers both the buffer size of each all-gather operator $o$, denoted as $\text{B}_{\text{ag}}(o)$, and its communication time \(T_c(\text{B}_{\text{ag}}(o))\). Parameters are prioritized based on the ratio \(\frac{T_c(\text{B}_{\text{ag}}(o))}{\text{B}_{\text{ag}}(o)}\), where a higher value indicates a greater reduction in communication time relative to memory cost. Since communication is less efficient for smaller messages, parameters with smaller buffer sizes are generally selected first.

\subsection{Adaptive offloading}\label{subsec:adaptive_offload}

Optimizer states such as momentum and variance used in the Adam optimizer~\cite{kingma2014adam} require a lot of GPU memory, but they are only needed for parameter updates and not during the forward or backward pass. Therefore, offloading them to CPU memory can significantly reduce peak memory usage. At the end of each backward pass, the optimizer states must be loaded back into GPU memory for parameter updates.
As implemented in ZeRO-3, the fully sharded approach is often combined with offloading, as both techniques help reduce memory requirements when training large-scale models.

\begin{algorithm}[t]
\caption{Adaptive offloading (forward)}\label{alg:offload}
\begin{algorithmic}
\REQUIRE Initial schedule $S_0$, Memory limit $M$, optimizer state fragments $OS = [os_{1}, os_{2}, \ldots]$

\STATE $S \gets []$ \COMMENT{List of operators in output schedule}
\STATE $M^- \gets 0$ \COMMENT{Offloaded size}
\STATE $M_{\text{peak}} \gets max_{o_i \in S_0} \text{P}_{\text{mem}}(o_i)$
\STATE $M_{opt} \gets \sum_{os_i \in OS} \text{B}_{\text{os}}(os_i)$ 
\STATE $OS_{\text{offload}} \gets []$

\FORALL{$os_i \in OS$}
  \IF{$M_{\text{peak}} + M_{\text{opt}} - \sum_{os_i \in OS_{\text{offload}}} \text{B}_{\text{os}}(os_i) > M$}
   \STATE $OS_{\text{offload}} \gets OS_{\text{offload}} \cup os_i$
   \STATE Append offload operator for $os_i$ to $S$
  \ENDIF
\ENDFOR
\FORALL{$o_i \in S_0$}
  \WHILE{$\text{P}_{\text{mem}}(o_i) + M_{\text{opt}} - M^- > M$}
    \STATE Pop $os_i$ from $OS_{\text{offload}}$
    \STATE Append operator to synchronize the copy of $os_i$ to $S$ and free memory of $os_i$
    \STATE $M^- \gets M^- + \text{B}_{\text{os}}(os_i)$
  \ENDWHILE
  \STATE Append $o_i$ to $S$
\ENDFOR

\RETURN $S$
\end{algorithmic}
\end{algorithm}

However, data transfers for offloading and reloading optimizer states can incur significant overhead. To address this challenge, we propose an approach called {\em adaptive offloading}, which improves upon existing offloading mechanisms in two ways. First, by monitoring actual memory usage during training, it minimizes the amount of data that needs to be offloaded. Only the portion of optimizer states that would exceed the memory limit is transferred to the host. Second, it takes advantage of the characteristic memory usage patterns in training. Since memory usage tends to increase during the forward pass and decrease during the backward pass, adaptive offloading schedules data transfers to overlap with computation. This overlap reduces the performance impact of moving data between GPU and CPU memory.

Algorithm~\ref{alg:offload} presents the pseudo-code for the forward pass with adaptive offloading of optimizer states. The optimizer states are divided into a large enough number of fragments. At the start of the forward pass, the algorithm schedules operators to initiate asynchronous offloading of fragments that exceed the memory limit. Since the copy operation is asynchronous, computation can run in parallel. The memory must only be freed after the copy operation completes. Therefore, the peak memory usage before each operator is checked based on profiling when scheduling. If the memory usage exceeds the limit, the algorithm adds operators to synchronize the copy and free memory for the fragments.
As memory usage typically increases during the forward pass, this strategy enables efficient overlap of computation and offloading.

We also iterate over the operators in the backward pass schedule, checking the available memory. If sufficient memory is available from that operator to the end, we schedule an operator to initiate asynchronous transfer the optimizer states back to the GPU. This approach enables overlapping the backward computation with the data transfer from host memory to GPU memory while adhering to memory constraints. In contrast to the forward pass, memory usage decreases during the backward pass, making it easier to hide reloading overhead.

\subsection{Composability}\label{subsec:composaibility}

\sysname organizes optimizations as modular compiler passes, enabling users to compose them flexibly based on their training goals and constraints. However, the effectiveness of these passes depends not only on their individual design but also on their application order and compatibility.

The optimization passes for proactive prefetching, selective unsharding, and adaptive offloading are designed to be applied after the pass that enables the fully sharded approach, which rewrites the computation graph to insert all-gather and release operations for sharded parameters. However, the order and combination of these passes can significantly affect overall performance.

A beneficial application order is to first apply proactive prefetching, followed by selective unsharding. This ordering ensures that selective unsharding can make use of the remaining memory after prefetching has opportunistically advanced communication while staying within memory constraints.
In contrast, applying selective unsharding before proactive prefetching can reduce the effectiveness of prefetching. Since unsharding attempts to utilize as much available memory as possible to retain parameters in memory, little capacity remains for prefetch buffers. As a result, the benefits of prefetching are largely lost.

Adaptive offloading is designed for scenarios where the model does not fit in GPU memory even after sharding. In practice, when adaptive offloading is applied, the memory available for proactive prefetching and selective unsharding is already limited, and the additional benefits from these passes tend to diminish.

\section{Evaluation}\label{sec:evaluation}

We conducted experiments to evaluate the performance of \sysname
across multiple dimensions, including computational efficiency, memory utilization, and correctness.

\begin{figure*}[t]
\begin{center}
\includegraphics[scale=0.3]{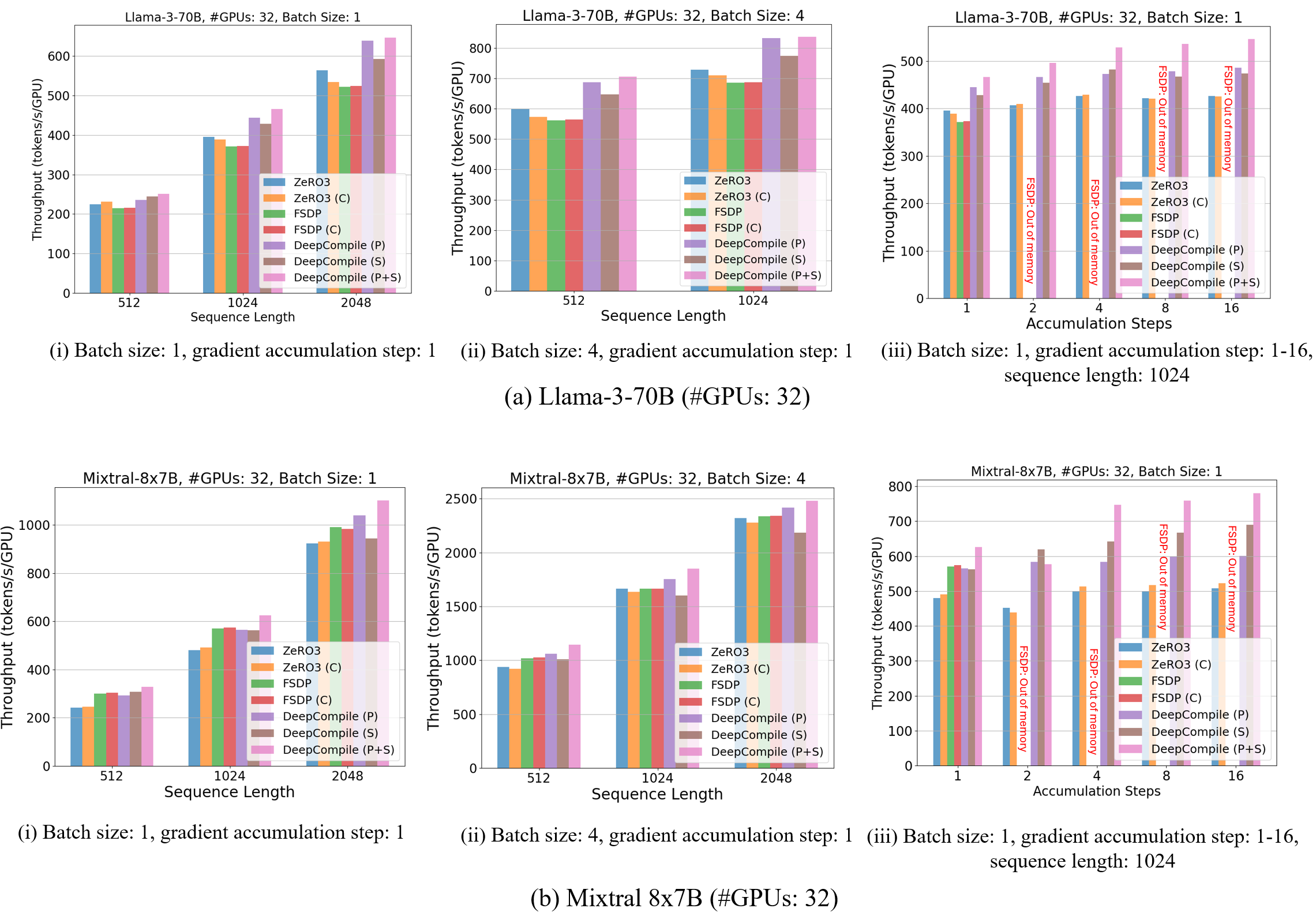}
\end{center}
\caption{Throughputs resulting from Llama-3 70B and Mixtral 8x7B models}
\label{fig:throughput}
\end{figure*}

\subsection{Experimental settings}

The experiments were performed on two or four servers, each with eight NVIDIA H100 GPUs (80GB per GPU), 1TB of system memory, and two Intel(R) Xeon(R) Platinum 8462Y+ CPUs. GPUs within each server were connected via NVLink, and servers were interconnected using InfiniBand (340GB/s observed bandwidth for allgather and reduce-scatter). Each server also included eight InfiniBand-connected HCAs.

We used Llama-3 70B and Mixtral 8x7B (47B) as representative dense and Mixture-of-Experts (MoE) models~\cite{jiang2024mixtral,shazeer2017sparsely,lepikhin2021gshard,fedus2022switch}. Both ran in bfloat16 with mixed precision~\cite{micikevicius2017mixed}, keeping FP32 copies of parameters, gradients, and Adam states. We applied activation checkpointing~\cite{chen2016training} at the Transformer layer level, recomputing each layer as a block.

We used PyTorch v2.6.0 and models from the HuggingFace Model Hub without any built-in parallelization support. The experiments were conducted with Python 3.10 and CUDA 12.4. ZeRO-3 (DeepSpeed v1.6.4) and PyTorch FSDP were used as the baseline frameworks, and \sysname was built on top of DeepSpeed. The optimization passes and profilers in \sysname were implemented in Python (approximately 2.6K lines), while the custom operators injected into the computational graph were implemented in C++ (approximately 900 lines).

Partial sharding strategies (e.g., sharding only optimizer states) are insufficient for our target models. Llama-3 70B and Mixtral 8x7B exceed single-GPU memory (140 GB and 94 GB), requiring all parameters, gradients, and optimizer states to be fully sharded. Thus, we use the fully sharded configurations of ZeRO-3 and FSDP in our evaluation.

Other large-scale training frameworks, such as Megatron-LM, are not included in our comparison. Although Megatron-LM has been widely adopted for large model training, it requires a specialized implementation where parallelism strategies are tightly coupled with model code. In contrast, we focus on model-agnostic approaches that can be applied to standard implementations. Similarly, frameworks that automate parallelization, such as Alpa~\cite{mo2022alpa}, FlexFlow~\cite{jia2019flexflow}, nnScalar~\cite{lin2024nnscaler}, and Unity~\cite{unger2022unity}, are not included, as they do not currently support the models used in this evaluation.

\subsection{Efficiency}

Figure~\ref{fig:throughput} presents the throughputs observed in our experiments.
We used Llama-3-70B and Mixtral-8x7B with 32 GPUs.
All throughput numbers reflect end-to-end time per training iteration, including the forward and backward passes as well as parameter updates.
We evaluated ZeRO-3 both without and with the PyTorch compiler enabled, labeled as \texttt{ZeRO3} and \texttt{ZeRO3 (C)}, respectively. Similarly, we include results for FSDP, labeled as \texttt{FSDP} and \texttt{FSDP (C)} depending on whether the PyTorch compiler is enabled.
For \sysname, we tested three configurations: enabling only proactive prefetching, enabling only selective unsharding, and enabling both in that order. These configurations are labeled as \texttt{\sysname (P)}, \texttt{\sysname (S)}, and \texttt{\sysname (P+S)}, respectively, in the charts.
Since the models fit within the available GPU memory at these scales, adaptive offloading was not applied in these experiments.

We conducted experiments with varying batch sizes, sequence lengths, and gradient accumulation steps. The per-GPU batch size (number of sequences) was set to 1, 2, or 4, and the sequence length to 512, 1024, or 2048. We also varied the gradient accumulation steps from 1 to 16. The effective batch size, measured in tokens, is given by:
$\text{\#Sequences per GPU} \times \text{Sequence length} \times \text{Accumulation steps} \times \text{\#GPUs}$.
Under these settings, the total number of tokens per step ranged from 8K to 4M, covering a broad range representative of common configurations used for training models at this scale.

For the system parameters defined in Table~\ref{tab:notation}, we set $M$ to 90\% of the available GPU memory to maintain a safety margin. $M_{\text{prefetch}}$ was set to 2~GB. Increasing this value allows \sysname to schedule more all-gather communications earlier, but we did not observe any clear performance benefit from setting it higher. The parameter $\alpha$ controls the aggressiveness of all-gather fusion. While larger values encourage more aggressive fusion, we found that increasing $\alpha$ beyond 1.5 had little effect on performance. Based on experiments with values ranging from 1.0 to 2.0, we fixed $\alpha$ to 1.5.
We used these values throughout all of the experiments.

\sysname consistently outperformed all baselines, with the largest gains observed when both proactive prefetching and selective unsharding were enabled (\texttt{\sysname (P+S)}).
With gradient accumulation set to 1, the smallest improvement over ZeRO-3 for Llama-3-70B (Fig~\ref{fig:throughput}(a),(i)(ii)) was 1.11×, observed at sequence length 512 and batch size 1. In this configuration, computation dominates communication, leaving limited room for our optimizations to hide communication latency via overlap. Consequently, the benefits of prefetching and unsharding are reduced.
Under all other conditions, \sysname achieved a 1.14×–1.18× speedup over ZeRO-3 and 1.16×–1.26× over FSDP.

For Mixtral-8x7B, the speedups over ZeRO-3 ranged from 1.07× to 1.35×, while those over FSDP ranged from 1.06× to 1.14× (Figure~\ref{fig:throughput}(b),(i)(ii)).
Mixtral-8x7B is an MoE model, where only a subset of parameters is activated for each input token. Due to its relatively small computation load compared to total parameter size, it suffers more from communication bottlenecks. As a result, similar to Llama-3-70B, the improvements were slightly larger when the batch size and sequence length were increased, enabling better overlap between communication and computation.

Figure~\ref{fig:throughput}(a)(iii) and (b)(iii) show throughput results when varying the gradient accumulation steps from 1 to 16. All numbers reflect the end-to-end time per training iteration, including all gradient accumulation steps.
As discussed in Section \ref{subsec:unsharding}, this setting significantly benefits from selective unsharding. In the Llama-3-70B experiment (Fig.~\ref{fig:throughput}(a)(iii)), \sysname achieves increasingly higher throughput as the accumulation step increases, with up to 1.28× improvement over ZeRO-3 at step 16.
In Mixtral-8x7B (Fig.~\ref{fig:throughput}(b)(iii)), the improvement is even more pronounced due to the model's heavier communication load. \sysname achieves up to 1.54× higher throughput compared to ZeRO-3 at step 16.

In contrast, FSDP fails to run with accumulation steps greater than 1, as it does not support accumulating gradients while keeping them partitioned. Instead, it gathers full gradients after each backward pass, which causes the total memory usage to exceed GPU capacity, leading to memory allocation failures.

\subsection{Memory Utilization}

\begin{figure}[t]
\begin{center}
\includegraphics[scale=0.3]{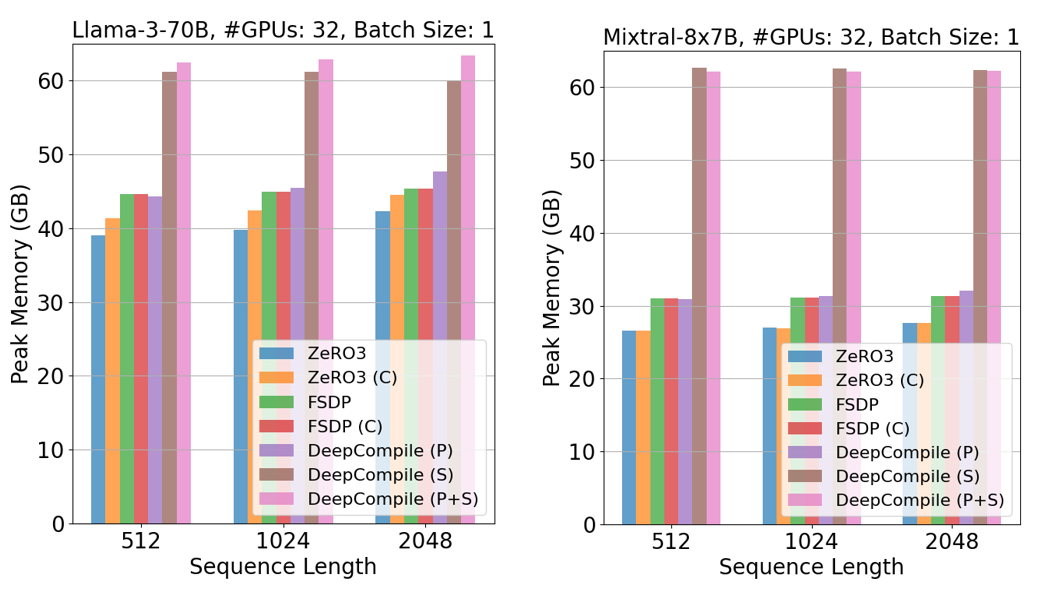}
\end{center}
\caption{Memory utilization from Llama-3 70B and Mixtral 8x7B models}
\label{fig:memory_util}
\end{figure}

The optimization passes described above allow \sysname to utilize GPU memory more effectively, contributing to increased throughput. We compare the memory utilization of \sysname against other baselines. Figure~\ref{fig:memory_util} presents the peak GPU memory usage for different models across various sequence lengths.

The memory footprints of ZeRO-3 and FSDP are approximately 40GB for Llama-3-70B and 30GB for Mixtral-8x7B. When selective unsharding is enabled, \sysname actively utilizes available memory to keep more parameters unsharded. We configure a safety margin by first reserving approximately 7GB for the CUDA driver, NCCL buffers, and other runtime components, and then applying an additional 10\% margin to the remaining memory. As a result, \texttt{\sysname (S)} and \texttt{\sysname (P+S)} consistently use around 65GB in all cases. This demonstrates that the selective unsharding mechanism dynamically adapts the amount of unsharded parameters based on the available memory at runtime.


\subsection{Adaptive Offloading}


The Llama-3-70B model requires at least 32 GPUs (each with 80GB memory) to fit entirely into GPU memory under our settings. However, by offloading optimizer states, we are able to run the model on just 16 GPUs. To evaluate the benefits of adaptive optimizer offloading (Section~\ref{subsec:adaptive_offload}), we compared the throughput achieved by DeepSpeed’s offloading mechanism combined with ZeRO-3 against our adaptive offloading approach.

\begin{figure}[t]
\begin{center}
\includegraphics[scale=0.27]{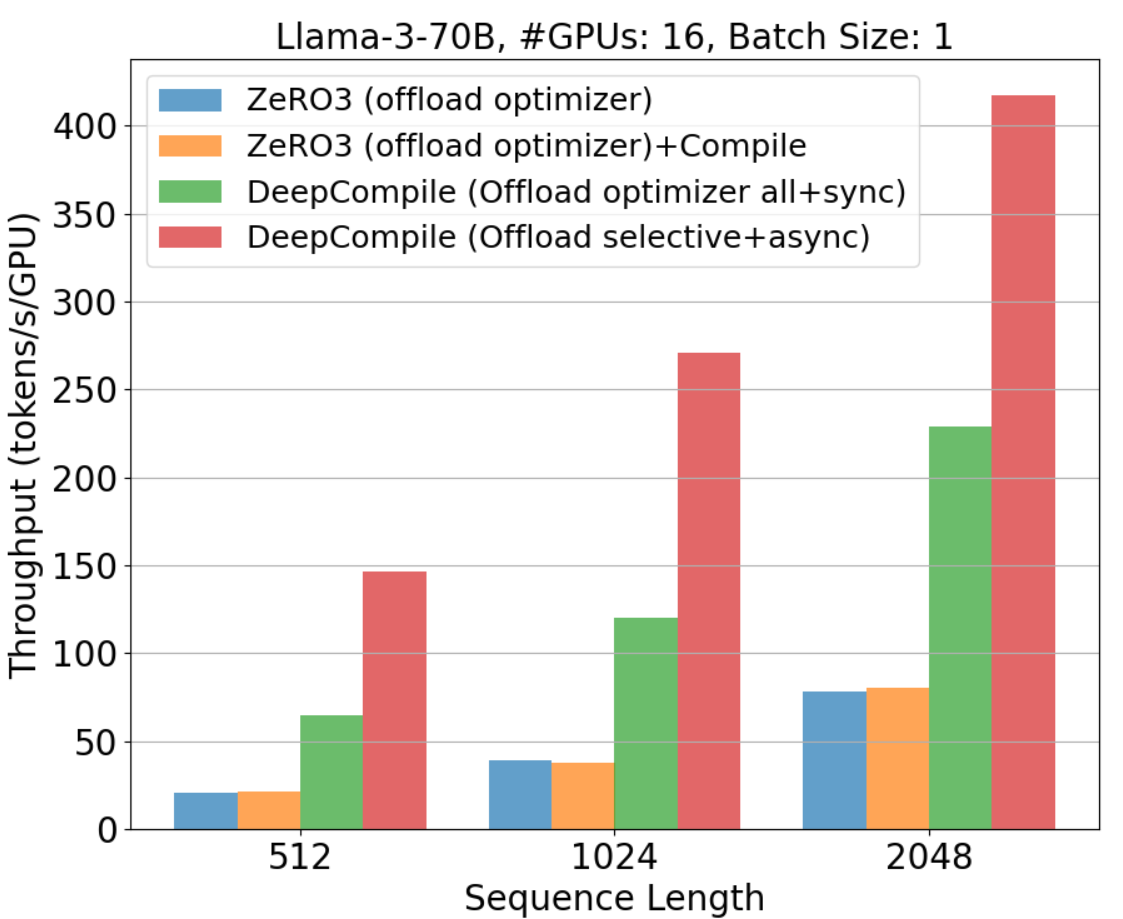}
\end{center}
\caption{Throughput results for adaptive offloading}
\label{fig:offload_eval}
\end{figure}

Figure~\ref{fig:offload_eval} presents the end-to-end iteration time for each method. \texttt{ZeRO3 (Offload optimizer)} and \texttt{ZeRO3 (Offload optimizer) + Compile} represent the results of using ZeRO-3 with and without the PyTorch compiler, respectively. Notably, ZeRO-3 not only offloads optimizer states to host memory but also performs parameter updates on the CPU. While this approach saves more memory than \sysname’s offloading, it significantly slows down parameter updates. Since this behavior makes it difficult to compare ZeRO-3's offloading directly with our method, which keeps parameter updates on the GPU, we also included results for \texttt{\sysname (Offload optimizer all + sync)}, which offloads all optimizer states at the beginning of the forward pass and reloads them synchronously at the end. Our adaptive offloading method is shown as \texttt{\sysname (Offload selective + async)}, which minimizes the amount of data offloaded and overlaps offloading and reloading asynchronously with computation.

The experimental results in Figure~\ref{fig:offload_eval} indicate that adaptive offloading (\texttt{\sysname (Offload selective + async)}) achieves up to 7.0× and 6.9× higher throughput than \texttt{ZeRO3 (Offload optimizer)} and \texttt{ZeRO3 (Offload optimizer) + Compile}, respectively, when the batch size is 1 and the sequence length is 1024. It also demonstrates a 2.6× improvement over \texttt{\sysname (Sync)}.

\subsection{Compilation time}

As shown in Fig.\ref{fig:opt_loop}, \sysname applies a sequence of optimization passes. In our experiments, we first executed the pass that inserts allgather and release operations. After a short warmup period (five training iterations in out experiments), we applied additional passes including proactive prefetching and selective unsharding.

These passes involve graph analysis and profiling, resulting in several minutes of overhead for our target models. Table~\ref{table:compile_time} reports the time required for each configuration, measured with batch size 1 and sequence length 512.

\begin{table}[h]
\caption{Time for compilation}
\label{table:compile_time}
\begin{tabular}{lp{2cm}p{1.7cm}l}
\hline
             & Proactive prefetching & Selective unsharding & Both    \\ \hline
Llama-3 70B  & 254.8 s               & 248.6 s          & 266.9 s \\ \hline
Mixtral 8x7B & 437.0 s               & 416.5 s          & 407.8 s \\ \hline
\end{tabular}
\end{table}

The variation in compile time across optimization configurations for the same model is less than 10\%, indicating that the additional cost of enabling proactive prefetching or selective unsharding is relatively minor compared to the base analysis and the pass to enable the fully-shareded approach. On the other hand, compile times vary more significantly across models, as seen in the longer durations for Mixtral-8x7B compared to Llama-3-70B. This difference stems from model-specific factors such as graph size and layer composition.

It is important to note that this compilation overhead occurs only once before training begins. During actual training, the compiled graph is reused without incurring any additional overhead. Given that model training typically consists of hundreds to thousands of iterations, the one-time compilation cost is negligible in practice.

Furthermore, the results of the optimization passes can be cached and reused across training runs as long as key settings such as batch size and sequence length remain unchanged. This reuse is particularly valuable when sweeping hyperparameters such as learning rates or dropout ratios, enabling users to avoid repeating the initial compilation and further reducing end-to-end overhead.

\subsection{Correctness}

\begin{figure}[t]
\begin{center}
\includegraphics[scale=0.45]{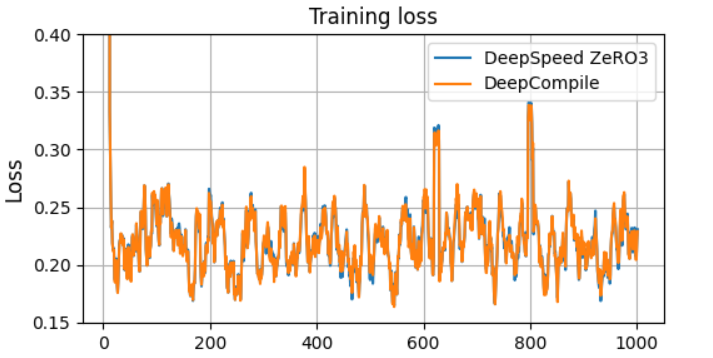}
\end{center}
\caption{Verification of loss values}
\label{fig:correctness}
\end{figure}

To verify the correctness of the optimization passes in \sysname, we compared the resulting loss values with those of ZeRO-3. We initialized Llama-3 70B model with random weights but the same random seed in both settings. We used the AG News corpus \cite{Zhang2015CharacterlevelCN} as training examples, with the sequence length and micro batch size set to 512 and 1, respectively. The learning rate was set to 1.5e-5. Figure~\ref{fig:correctness} shows the training losses for both settings. Although some operators are non-deterministic and introduce subtle differences, the loss curves were closely aligned.

\begin{table*}[t]
\caption{Comparison of \sysname with existing distributed training systems}
\label{tab:comparison}
\centering
\begin{tabular}{lcccc}
\toprule
\textbf{Feature} & \textbf{ZeRO-3 / FSDP~\cite{rajbhandari2020deepspeed, fsdp_paper}} & \textbf{SimpleFSDP~\cite{zhang2024simplefsdpsimplerfullysharded}} & \textbf{Alpa / Unity~\cite{mo2022alpa,unger2022unity}} & \textbf{\sysname} (ours)\\
\midrule
No code modification               & \checkmark & \checkmark & \xmark & \checkmark \\
Compiler-level optimization           & \xmark     & \checkmark & \checkmark & \checkmark \\
Profiling guided optimization      & \xmark     & \checkmark & \checkmark & \checkmark \\
Communication overlap              & \checkmark & \checkmark & \xmark & \checkmark \\
Composable optimizations           & \xmark     & \xmark     & \checkmark & \checkmark \\
Memory-aware scheduling            & \xmark     & \xmark & \xmark & \checkmark \\
\bottomrule
\end{tabular}
\end{table*}

\section{Related Work}
\label{sec:related_work}

Distributed training has become a central topic in deep learning system design, leading to a range of frameworks that support model parallelism, memory optimization, and scaling strategies. We summarize key systems and contrast their capabilities with our approach in Table~\ref{tab:comparison}.

\vspace{1mm}
\noindent\textbf{Framework-level sharding}  
Fully sharded training frameworks such as ZeRO-3~\cite{rajbhandari2020deepspeed} and PyTorch FSDP~\cite{fsdp_paper} partition model parameters and optimizer states across devices and apply runtime optimizations like prefetching and unsharding. These optimizations are typically implemented using hooks inserted at the framework level and rely on heuristics that do not leverage global analysis or profiling feedback. As a result, they offer limited control over the scheduling of communication and memory operations, and cannot coordinate multiple interacting optimizations in a globally informed manner.

\vspace{1mm}
\noindent\textbf{Compiler-based prefetching}  
SimpleFSDP~\cite{zhang2024simplefsdpsimplerfullysharded} uses a compiler to improve communication scheduling for fully sharded training. It analyzes operator-level memory usage to decide whether to fuse multiple all-gather operations. However, it focuses solely on prefetching and does not support coordination across multiple optimizations or feedback loops between passes. In addition, while it considers memory usage for fusion decisions, it does not take memory dynamics into account when determining the placement of all-gather operations. 
\sysname generalizes this idea by introducing a pass-based architecture in which each optimization pass can access updated profiling results and respond to memory dynamics.

\vspace{1mm}
\noindent\textbf{Automatic parallelization planning}
FlexFlow~\cite{jia2019flexflow} proposed selecting optimal combinations of data, tensor, and pipeline parallelism based on cost models. Building on this idea, recent systems such as Alpa~\cite{mo2022alpa}, nnScaler~\cite{lin2024nnscaler}, and Unity~\cite{unger2022unity} integrate parallelism planning with compiler-based optimization techniques. These systems focus on cost-model-driven planning and static graph analysis to combine parallelism strategies and GPU-local optimizations. However, they do not account for runtime memory usage trends or support techniques such as communication–computation overlap and memory-aware offloading.

\sysname complements these approaches by emphasizing profiling-guided, memory-centric optimization using modular passes that adapt to runtime behavior. Although our current focus is on the fully sharded approach, \sysname can also express other parallelism strategies. For example, tensor parallelism can be realized by identifying Transformer blocks and injecting all-reduce operations; pipeline parallelism can be enabled by partitioning the graph into stages with balanced compute and memory load; and sequence parallelism~\cite{jacobs2023deepspeedulyssesoptimizationsenabling,liu2023ringattentionblockwisetransformers} can be supported through all-to-all or peer-to-peer communication.

A key distinction is that \sysname's modular pass framework enables coordinated composition of such parallelization strategies with memory-aware techniques like offloading. This interaction has received little attention in prior work and represents a promising direction for future research.

\section{Conclusion}

We presented \sysname, a compiler-based optimization framework for distributed deep training. By transforming user-defined models into computation graphs and applying a series of profiling-guided optimization passes, \sysname enables fine-grained memory-aware scheduling and communication planning across passes. This approach contrasts with existing systems like ZeRO-3 and FSDP, which lack flexibility in communication timing and memory management.

Built on a fully sharded foundation, \sysname incorporates three key optimization passes: proactive prefetching to improve communication–computation overlap, selective unsharding to reduce communication, and adaptive offloading to manage memory when resources are limited. Each pass utilizes updated profiling feedback, allowing the system to compose  optimizations based on their combined effects.

Our evaluation demonstrates that \sysname consistently outperforms existing baselines. It achieves up to 1.28× and 1.54× throughput improvements on Llama 3 70B and Mixtral 8×7B models, respectively. In memory-constrained settings, adaptive offloading enables up to 7.01× throughput gains by overlapping data transfers with computation.

\sysname represents a step forward in compiler-driven distributed training. Its graph transformation capabilities and runtime-aware optimization passes provide a scalable and automated path to efficient parallelization. As discussed in Section~\ref{sec:related_work}, future work will extend \sysname to support a broader class of optimizations, including communication and memory scheduling.

\bibliographystyle{ACM-Reference-Format}
\bibliography{deepcompile}

\end{document}